\documentclass{rnaastex}
\usepackage{url}

\begin{document}

\title{On the Consequences of the Detection of an Interstellar Asteroid}

\correspondingauthor{Gregory Laughlin}
\email{gregory.laughlin@yale.edu, kbatygin@caltech.com}

\author[0000-0002-3253-2621]{Gregory Laughlin}
\affiliation{Department of Astronomy, Yale University \\New Haven, CT 06511, USA}

\author{Konstantin Batygin}
\affiliation{Division of Geological and Planetary Sciences, California Institute of Technology\\ Pasadena, CA 91125, USA}


\section{} 

The arrival of the robustly hyperbolic asteroid A/2017 U1  \citep{mpec2017a} seems fortuitously timed to coincide with the revival of the \textit{AAS Research Notes} \citep{vishniac2017}. Both the sparse facts surrounding A/2017 U1's properties and trajectory, as well as its apparently startling ramifications for the planet-formation process, are readily summarized in less than a thousand words.

With an eccentricity, $e=1.2$, A/2017 U1 had a pre-encounter velocity, $v_{\infty} = 26\,{\rm km\,s^{-1}}$ relative to the solar motion, and its direction of arrival from near the solar apex is entirely consistent with Population I disk kinematics \citep{mamajek2017}. At periastron, A/2017 U1 passed $q=0.25$~AU from the Sun, momentarily reaching a heliocentric velocity of 88 ${\rm km\,s^{-1}}$. Despite briefly achieving solar irradiation levels $I>20\,{\rm kW\,m^{-2}}$, deep images produced no sign of a coma \citep{mpec2017b}, suggesting that the object has a non-volatile near-surface composition. Its spectrum, moreover, shows no significant absorption features and is considerably skewed to the red \citep{2017arXiv171009977M}. The object's $H=22$ absolute magnitude, coupled with albedo assumed to be of order $A\sim0.1$ implies that it has a diameter $d\sim160$ meters.

Adopting an assumed solar encounter rate\footnote{With a \textit{single} occurrence, this once-per-ten-years encounter rate is tied to the time that Pan-STARRS \citep{2016arXiv161205560C} and similar surveys have been operating, and is, of course, highly uncertain.}, $\Gamma=n\sigma v_{\odot} \sim 0.1\,{\rm yr^{-1}}$, a detection/interaction cross section, $\sigma \sim 3\,AU^{2}$ (aided by gravitational focusing), and a velocity of the Sun relative to the Population-I standard of rest, $v_{\odot}=20 \,{\rm km\,s^{-1}}$, one estimates the space density of objects similar to A/2017 U1 to be of order $n\sim1/100\,{\rm AU}^{-3}$. This implies a total of $N_{\rm tot}\sim2\times10^{26}$ such objects in the galaxy (assuming a cylindrical galaxy with $R=3\times10^{4}\,{\rm pc}$ and $H=10^{3}\, {\rm pc}$). If we assume a density $\rho=1\,{\rm g\,cm^{-3}}$ (characteristic of similarly-sized Kuiper belt objects), the implied galactic mass in such bodies is $M_{\rm tot}\sim10^{11}\,M_{\oplus}$.

While extrapolations from a single event bring obvious attendant uncertainties, the A/2017 U1 encounter implies of order an Earth mass of ejected planetesimals for every star in the galaxy. The lack of a detectable coma during the A/2017 U1 flyby is additionally puzzling, reminiscent of the rare solar system Damocloids \citep{2005AJ....129..530J}, or, alternately, refractory-composition asteroids, rather than Oort-cloud comets.

Our own solar system has contributed many volatile-rich planetesimals to the galaxy. Specifically, within the framework of the so-called Nice model of early solar system evolution, \citep{2005Natur.435..459T, 2008Icar..196..258L}, a transient period of dynamical instability is triggered in response to interactions between the giant planets and a primordial disk comprising $\sim30 M_{\oplus}$. In numerical realizations, nearly all of this material is expelled into the interstellar medium as the instability unfolds, leaving behind today's severely mass-depleted Kuiper belt. Given the universality of \textit{N}-body evolution, one can speculate that similar sequences of events are a common feature of planetary system evolution. 

The efficient ejection of planetesimals requires a massive secondary body with specific orbital conditions. For a planet with mass, $M_{\rm p}$, radius, $R_{\rm p}$ and semi-major axis, $a$ to be capable of efficiently ejecting planetesimals from a planetary system through the action of gravity assist \citep{Bond1996}, the ratio, $f=v_{o}/v_{c}=((M_{\rm p}a)/(M_{\star}R_{\rm p}))^{1/2}$, must significantly exceed unity. In our solar system, all four giant planets can induce ejections. Jupiter, with $f_{\rm J}=3.2$, and Neptune, with $f_{\rm N}=3.1$ are more capable than the others.

Among the known extrasolar planets (Figure 1), neither the hot Jupiters, nor the far more numerically dominant population of super-Earths -- which typically have $M\sim10 M_{\oplus}$, $R\sim3 M_{\oplus}$, and $a\sim 0.2$  \citep{2015ARA&A..53..409W} -- can eject planetesimals. Approximately 5-10\% of solar-type stars (but not M-dwarfs) \textit{do} harbor massive intermediate-period planets with $f>1$ \citep{2008PASP..120..531C, 2016ApJ...817..104R} that could potentially eject primarily refractory material from their systems. If A/2017 U1 points to a population bereft of volatiles, each such planet would be required to eject $>$100 $M_{\oplus}$ in shards of rock and metal, a quantity that seems challenging to generate. In all likelihood, the comparatively low occurrence rate of these planets allows them only minor contributions to the galaxy's burden of free-floating planetesimals, implying that A/2017 U1 contains a substantial volatile component despite its lack of coma. The primary burden of generating interstellar asteroids likely falls on an as-yet unobserved population of long-period sub-Jovian planets.

Recent analyses of the radius-mass relationship of extrasolar planets detected by \textit{Kepler} have found essentially linear proportionality between the two quantities. Specifically, \citet{{2013ApJ...772...74W}} find $M_{\rm p}\propto R_{\rm p}$ for planets less massive than $M_{\rm p} <20M_{\oplus}$, while \citet{2013ApJ...768...14W} obtain $M_{\rm p} \propto R_{\rm p}^{0.93}$ as the best fit for objects with $1.5 R_{\oplus} < R_{\rm p} < 4 R_{\oplus}$. Crucially, a linear radius-mass relationship implies an escape velocity that is independent of the planetary characteristics. In turn, this means that among sub-Jovian extrasolar planets, the ratio $f$ is largely a function of the stellar mass and orbital radius.

Upon substitution of the relevant constants, we find that for solar-mass stars, the characteristic semi-major axis -- beyond which ejection of planetesimals is readily accomplished by relatively low-mass planets -- lies at $a\sim5\,$AU. (This ``throw line'' diminishes to $a\sim1\,$AU for $M_{\star}=0.2M_{\odot}$ M-dwarfs). Coincidentally, these values roughly correspond to the ice-sublimation lines of the respective stars, and strongly hint at the ubiquity of sub-Jovian planets residing at stellocentric radii of order a few astronomical units.

\acknowledgements
We thank Joel Ong for an insightful comment and Mike Brown for a useful conversation.

\begin{figure}[h!]
\begin{center}
\includegraphics[scale=0.75,angle=0]{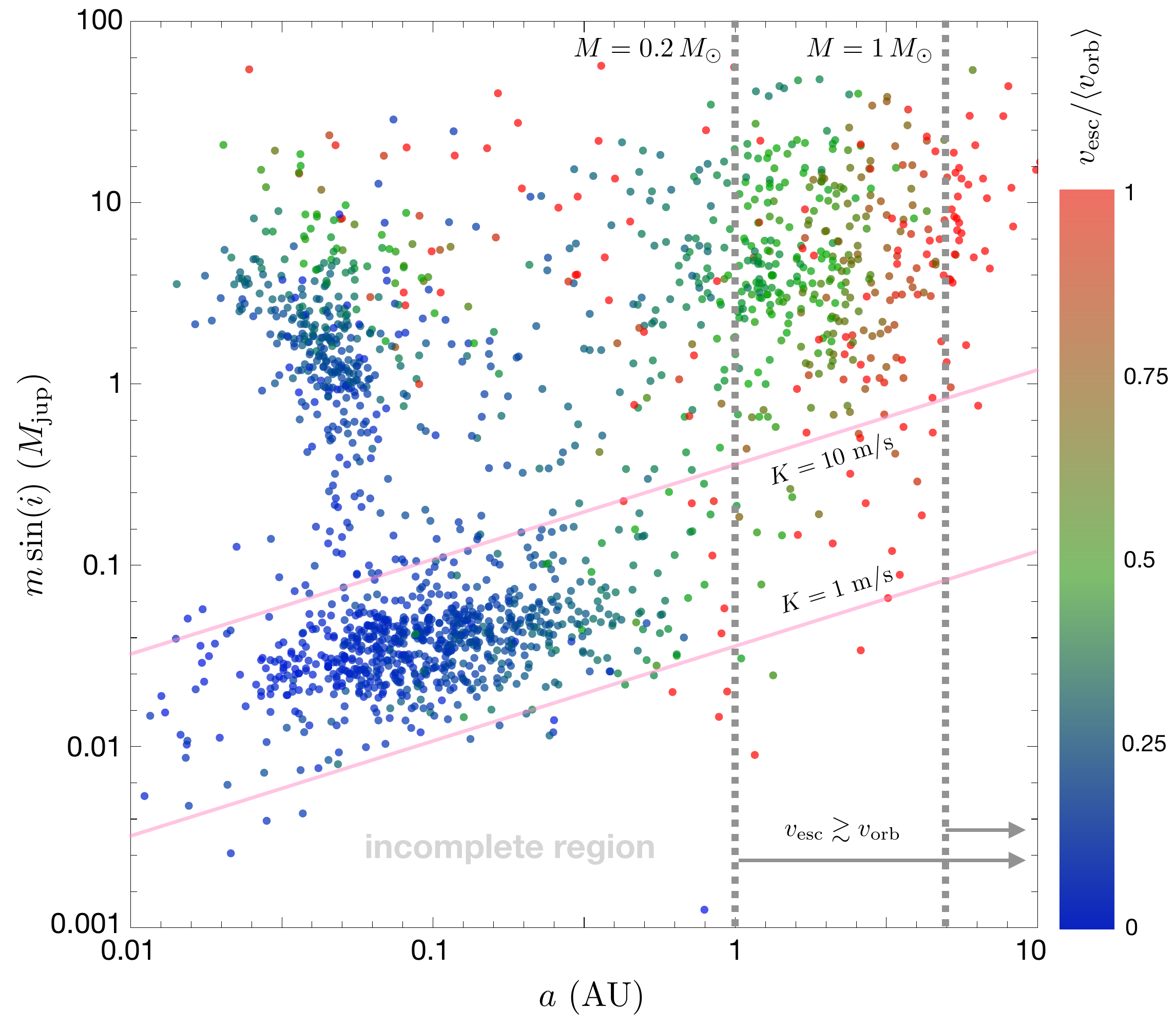}
\caption{Confirmed extrasolar planets with physical properties drawn from \url{https://exoplanetarchive.ipac.caltech.edu/}. Planetary (minimum) mass is shown as a function of the semi-major axis, and the data points are color-coded according to the estimated ratio of the escape velocity and the circular orbital velocity. For objects without direct mass or radius measurements, these quantities are inferred using the linear mass-radius relationship of \citet{2013ApJ...772...74W}, with a ceiling of Jupiter radius. In addition to the observations, radial velocity half-amplitudes of $K=1\,$m/s and $K=10\,$m/s are shown for $M=1M_{\odot}$. The interstellar asteroid A/2017 U1 implies a vast and cool, as-yet undetected population of planets with $f=v_{\rm esc}/v_{\rm orb}\gtrsim1$. \label{fig:1}}
\end{center}
\end{figure}

\end{document}